\documentclass[aps,reprint,amsmath,amssymb,graphicx,longbibliography]{revtex4-2}

\newcommand{\Tc}{T_{\textrm c}}

\newcommand{\kB}{k_{\rm B}}

\newcommand{\ac}{a_{\textrm c}}

\usepackage{float}
\usepackage{bm}
\usepackage{fourier}
\usepackage{amsmath,amssymb}
\usepackage{mathtools}

\usepackage{times}
\usepackage{upgreek}
\usepackage{psfrag} 
\usepackage{latexsym} 
\usepackage{amstext}
\usepackage{amsxtra} 

\usepackage{dcolumn}

\usepackage{textcomp}
\usepackage{amsfonts}
\usepackage{graphicx}
\usepackage{bm}
\usepackage{color}
\usepackage{braket}
\usepackage{units}
\usepackage[svgnames]{xcolor}

\definecolor{myColor}{rgb}{0.02,0.12,0.3}
\definecolor{myciteColor}{rgb}{0.39,0.7,0.89}
\usepackage[colorlinks=true,citecolor=myColor,linkcolor=myColor,urlcolor=myColor]{hyperref}

\def\be{\begin{equation}}
\def\ee{\end{equation}}

\def\nobreakbefore{%
  \relax\ifvmode\else
    \ifhmode
      \ifdim\lastskip > 0pt\relax
        \unskip\nobreakspace
      \else 
        \nobreakspace
      \fi
    \fi
  \fi
}
\let\oldcite\cite
\renewcommand\cite{\nobreakbefore\oldcite}

\begin{document}

\title{
Interacting Bose-condensed gases\\
{\normalfont \small \textit{Invited Contribution to Encyclopedia of Condensed Matter Physics, 2nd edition}}}

\author{Christoph Eigen} 
\affiliation{Cavendish Laboratory, University of Cambridge, J. J. Thomson Avenue, Cambridge CB3 0HE, United Kingdom}
\author{Robert P. Smith}
\affiliation{Clarendon Laboratory, University of Oxford, Parks Road, Oxford OX1 3PU, United Kingdom}

\date{\today{}}

\begin{abstract}
We provide an overview of the effects of interactions in Bose-condensed gases.
We focus on phenomena that have been explored in ultracold atom experiments, covering both tuneable contact interactions and dipolar interactions. Our discussion includes: modifications to the ground state and excitation spectrum, critical behaviour near the Bose--Einstein condensation temperature, the unitary regime where the interactions are as strong as allowed by quantum mechanics, quantum droplets in mixtures, and supersolids in dipolar gases.
\end{abstract}

\maketitle


\section{Key Objectives}
\begin{itemize}
    \item Summarize ideal-gas Bose--Einstein condensation. 
    \item Review the consequences of weak repulsive contact interactions on the ground state, excitations, and thermodynamics of Bose condensed gases.
    \item Showcase recent work on the unitary Bose gas.
    \item Highlight novel effects in condensates that experience attractive or dipolar interactions, including the formation of quantum droplets and supersolids.  
    
\end{itemize}

\section{Introduction} 
\label{intro}

Bose--Einstein condensation (BEC) plays an important role across a variety of quantum systems and phenomena, from superconductivity \cite{Onnes:1911}, and exciton \cite{Deng:2002,Deng:2006, Kasprzak:2006,Balili:2007} and magnon~\cite{Demokritov:2006} condensation in the solid state, to superfluidity in liquid helium \cite{Kapitza:1938, Allen:1938}, to photon condensates in optically pumped dye-filled mircocavities \cite{Klaers:2010}, and to its most direct demonstration in ultracold dilute gases~\cite{Anderson:1995,Davis:1995a}. 
A unique aspect of the Bose condensed state is that the (phase) order it displays is not primarily the result of interparticle interactions, but can occur in an ideal (noninteracting) gas solely due to quantum statistics.

In this article we discuss how the interplay of interparticle interactions with the underlying phase order affects the Bose-condensed state, and the host of fascinating phenomena that ensue.
While such questions first arose in the context of liquid helium (a strongly interacting fluid), here we focus our discussion on dilute ultracold atomic gases. In these systems the interaction strength can readily be tuned, allowing controlled access to both subtle and dramatic interaction effects. This enables tests of fundamental many-body theories and the progressive exploration towards strong interactions, where a first principles description is intractable.

We further restrict our discussion to continuous Bose gases in three dimensions (3D), leaving aside fermionic systems where BEC can occur as a consequence of pairing~\cite{Inguscio:2007,Zwerger:2011,Zwierlein:2014}, Bose gases confined in optical lattices~\cite{Bloch:2005, Morsch:2006}, and lower-dimensional systems where the enhanced role of fluctuations often precludes BEC~\cite{Cazalilla:2011,Hadzibabic:2011}.
Note that we provide only a brief overview here; many of the topics are covered in much greater detail elsewhere~(see \emph{e.g.} \cite{Pethick:2002, Pitaevskii:2016}).

The article is organised as follows. In Section \ref{sec:idealBEC} we briefly review Bose--Einstein condensation in a gas of noninteracting particles, before exploring the effect of different interparticle interactions on the Bose condensed state: Sections \ref{sec:WeakRepulsiveContact}-\ref{sec:Unitary} consider repulsive contact interactions in, respectively, the weak and strong limit. Section \ref{sec:AttractiveAndDipolar} explores phenomena which occur due to either attractive or dipole-dipole interactions. Finally, we conclude and provide a brief outlook in Section \ref{sec:conclusions}.

\section{Noninteracting Bose gases} 
\label{sec:idealBEC}
BEC is a phase transition associated with the onset of long-range phase order, which results from the macroscopic occupation of a single-particle state. 
This is a consequence of Bose--Einstein statistics and occurs at a critical temperature $\Tc$ far above the single-particle energy-level spacing. 

\begin{figure}[b!]
\centerline{\includegraphics[width=\columnwidth]{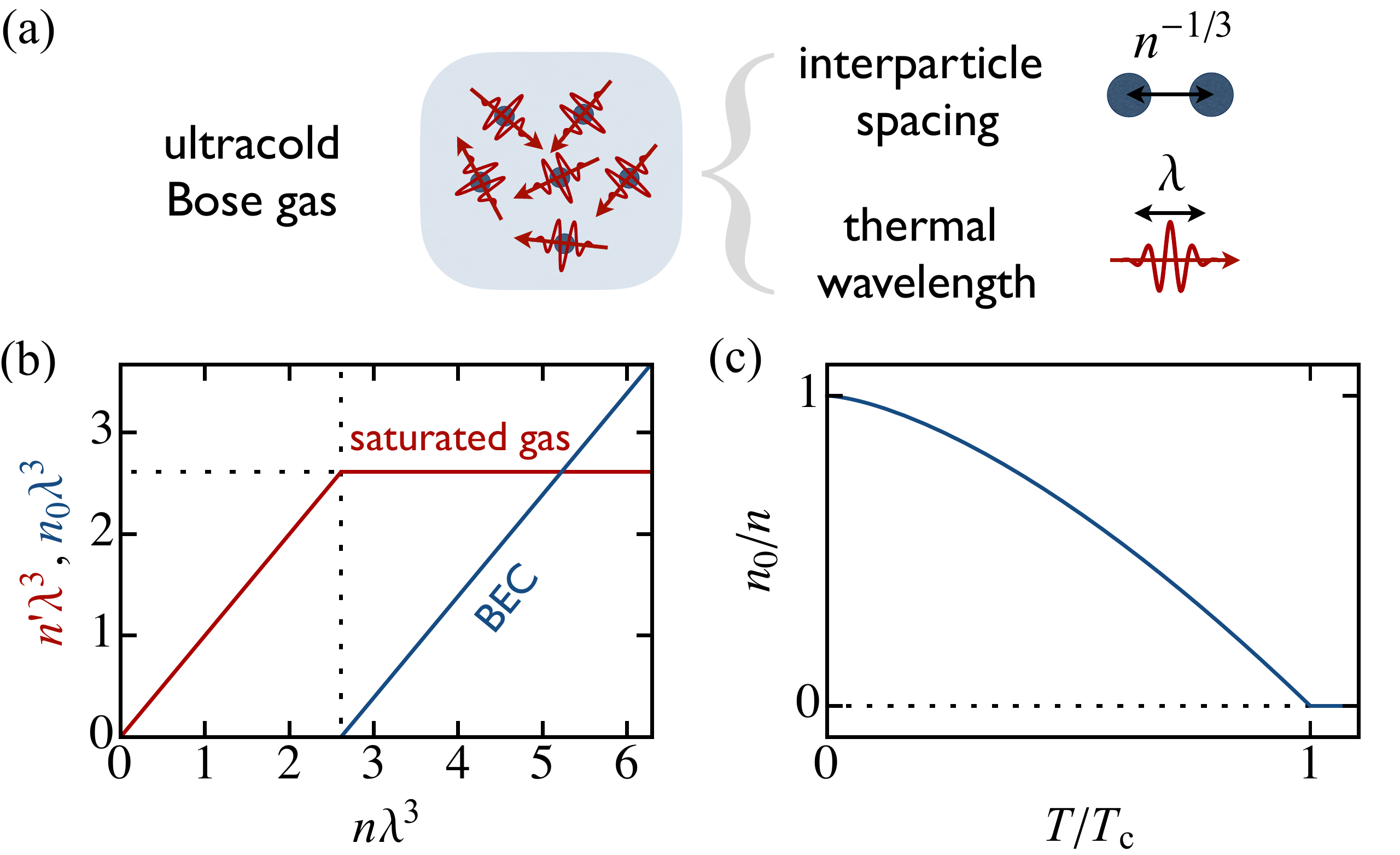}}
\caption{\textbf{Ideal Bose gases.}
(a) Noninteracting ultracold Bose gases feature two relevant lengthscales: the interparticle spacing $n^{-1/3}$ and the thermal wavelength $\lambda$. (b) Excited- and ground-state phase space densities versus total phase space density.
Above a critical $\mathcal{D}_{\textrm c}$ (dotted vertical line), $n'\lambda^3$ saturates and $n_0\lambda^3$ grows.
(c) Condensed fraction $n_0/n$ versus $T/\Tc$ [Eq.~(\ref{eq:uniformCF})].
}
\label{fig:idealgas}
\end{figure}

Qualitatively, the transition can be understood by considering lengthscales in a quantum gas [see Fig.~\ref{fig:idealgas}(a)]. The two relevant lengthscales are the interparticle spacing, $n^{-1/3}$, set by the gas density $n$, and the (quantum-mechanical) thermal wavelength $\lambda \; = \; [2\pi\hbar^2 /(m \kB T)]^{1/2}$ set by the temperature $T$ and the mass $m$ of the gas particles. Condensation occurs when the thermal wavelength is large enough to be comparable to the interparticle spacing ($\lambda \gtrsim n^{-{1/3}}$).

More quantitatively, we consider the occupation of a momentum state $\bf p$ in a homogeneous ideal (noninteracting) gas, which is given by the Bose distribution
\begin{equation}\label{eq:bosep}
    f_{\rm B}=\frac{1}{\mathrm{e}^{(\varepsilon-\mu)/(\kB T)}-1} \; , 
\end{equation}
where $\mu$ is the chemical potential and $\varepsilon=p^2/(2m)$ the energy of the state.
The total number of particles in the gas is found by summing over all possible states:
\begin{equation}\label{smith:eq:sum}
    N=\sum_{\varepsilon}\frac{g_{\varepsilon}}{\mathrm{e}^{(\varepsilon-\mu)/(\kB T)}-1} \; ,
\end{equation}
where the degeneracy factor $g_{\varepsilon}$ accounts for the number of states with a given $\varepsilon$. Demanding all terms in the sum to be real positive numbers requires $\mu \leq \varepsilon_0$, where $\varepsilon_0$ is the ground-state energy. As $\mu$ approaches $\varepsilon_0$ from below (practically achieved by \emph{e.g.}~increasing $N$) the ground-state occupation can become arbitrarily large [see Eq.~(\ref{eq:bosep})], whereas the occupation of all the other states (the excited states) sums to a finite (critical) number in 3D. 
This statistical saturation of excited states [graphically demonstrated in Fig.~\ref{fig:idealgas}(b)] is, in essence, the mechanism for BEC.
At fixed temperature, as particles are added to a Bose gas they populate the excited states until their population saturates at the critical atom number. From this point on, any additional particles must enter the ground state, which thus becomes macroscopically occupied.

In the thermodynamic limit, in which both $N$ and the volume of the system $V$ are large, while the particle density $n$ is finite, we may (semiclassically) approximate the sum over excited states by an integral. The density of excited states $n'$ (also known as the thermal density), is given by:
\begin{equation}\label{smith:eq:densityint}
    n'=\int\frac{\mathrm{d}\mathbf{p}}{(2\pi \hbar)^3}\frac{1}{\mathrm{e}^{(p^2/(2m)-\mu)/(\kB T)}-1}=\frac{g_{3/2}(\mathrm{e}^{\mu/(\kB T)})}{\lambda^3} \; ,
\end{equation}
where the polylogarithm $g_{3/2}(x)=\sum_{l=1}^{\infty}x^{l}/l^{3/2}$. We can re-express this result in terms of the phase space density
$n'\lambda^3=g_{3/2}(\mathrm{e}^{\mu/(\kB T)})$,
which reaches a maximum (critical) value when $\mu=0$ (as $\varepsilon_0\rightarrow 0$), such that
$n'\lambda^3=\mathcal{D}_{\textrm c}=g_{3/2}(1)=\zeta(3/2)\approx2.612$, where $\zeta$ is the Riemann function.

For a fixed $n$, the BEC transition temperature is given by
\begin{equation}\label{eq:uniformTc}
   \kB \Tc=\frac{2\pi \hbar^2}{m}\left(\frac{n}{\zeta(3/2)}\right)^{2/3} \; ,
\end{equation}
and the condensed fraction
\begin{equation}\label{eq:uniformCF}
   n_0/n=1-(T/\Tc)^{3/2} \; ,
\end{equation}
where $n_0$ is the condensate density [see Fig. \ref{fig:idealgas}(c)].

In the following section we discuss how this picture changes in the presence of weak repulsive `contact' interactions, a regime readily accessible in ultracold-gas experiments.

\section{Weak repulsive contact interactions} 
\label{sec:WeakRepulsiveContact}
 
For a dilute atomic gas the effective low-energy interaction between two atoms at $\mathbf{r}$ and $\mathbf{r}'$ can be approximated as a contact interaction $g \delta(\mathbf{r}-\mathbf{r}')$ with $g=4\pi\hbar^2 a/m$, where $a$ is the $s$-wave scattering length.  The interactions are thus quantified by a single lengthscale $a$, which can be tuned experimentally using so-called Feshbach resonances \cite{Chin:2010}.

The consequences of interactions are often grouped into mean-field (MF) and beyond-mean-field (BMF) effects. 

Simplistically speaking, MF effects result from the effective potential felt by a particle due to the average local particle density. Within the Hartree--Fock approximation \cite{Dalfovo:1999} the excited-state atoms feel an effective potential $V'_{\rm int}=g(2n_0+2n')$ and, owing to the lack of the exchange interaction for particles in the same state, the ground-state atoms experience a different interaction potential $V_{\rm int, 0}=g(n_0+2n')$.
In a homogeneous system, these interaction potentials lead to a uniform energy offset, \emph {i.e.}~ $\varepsilon= p^2/2m+V_{\rm int}$ in Eq.~(\ref{eq:bosep}). This is accompanied by a simple shift of the chemical potential such that it is only the factor of two difference between the condensate-condensate and condensate-thermal interaction potentials that leads to non-trivial interaction effects at the MF level.
By contrast, in inhomogeneous systems the entire interaction potential is important as it effectively  modifies the form of the trap. This often results in MF effects being more prominent in inhomogeneous systems; see \cite{Smith:2017} for a detailed discussion.

Note that, at the mean-field level, the condensate behaviour can be described by a macroscopic wavefunction that obeys the Gross--Pitaevski equation (see \cite{Pitaevskii:2016}), 
a nonlinear Schr\"odinger equation.

Beyond-mean-field effects arise due to changes of the many-body wavefunction, sometimes referred to as quantum fluctuations.
These effects become significant if the interaction energy $\sim gn$ is no longer negligible compared to the relevant (kinetic) energy. In particular, significant effects occur whenever low-energy (long-wavelength) excitations play a central role.

We now consider the effect of interactions on the elementary excitations of a Bose-condensed system, on its ground state, and on its behavior near $\Tc$.

\subsection{Excitations and sound}

In a homogeneous system, at the MF-level the excitation spectrum is simply given by  $\epsilon(k)=p^2/(2m)+V'_{\rm int}-V_{\rm int,0}=\hbar^2 k^2/(2m)+gn_0$, where $k=p/\hbar$ is the excitation wavenumber.

\begin{figure}[b!]
\centering
\includegraphics[width=1\columnwidth]{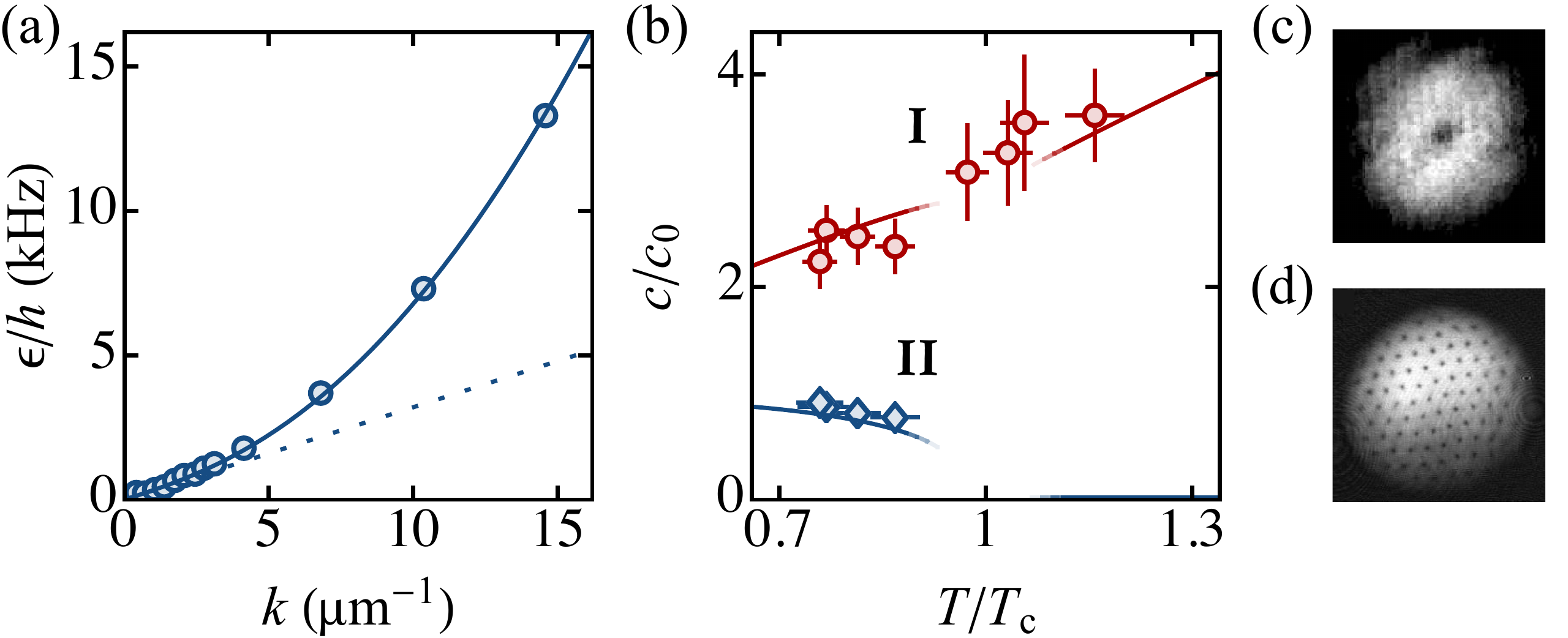}
\caption{\textbf{Excitations in interacting degenerate Bose gases.} (a) Excitation spectrum of a weakly interacting Bose--Einstein condensate, measured in a harmonic trap~\cite{Steinhauer:2002}. The solid line shows the fitted Bogoliubov form, with a linear spectrum at low momenta (dotted line), the slope of which gives the speed of sound.
(b) 
At finite temperature a Bose fluid supports two sound modes, here measured in a box-trapped ultracold gas~\cite{Hilker:2022}; the speeds are normalized by the Bogoliubov speed and the solid lines show the two-fluid model predictions.
(c) Vortices are another important type of excitation; the image shows a single vortex generated via stirring a Bose--Einstein condensate with a laser beam~\cite{Madison:2000a}. 
(d) When large numbers of vortices are generated the interactions between them lead to the formation of an Abrikosov lattice~\cite{AboShaeer:2001}.
Panels adapted from: (a)\cite{Steinhauer:2002}, (b)\cite{Hilker:2022}, (c)\cite{Madison:2000a}, (d)\cite{AboShaeer:2001}.
}
\label{fig:Sound}
\end{figure}

For low-energy (low-$k$) excitations this MF approach is no longer adequate; to next order the effect of interactions can be calculated using the Bogoliubov transformation\cite{Bogoliubov:1947}, which enables a description in terms of noninteracting quasi-particles; this approach predicts an excitation spectrum
\begin{equation}
    \epsilon(k)=\sqrt{\frac{\hbar^2 k^2}{2m}\left(\frac{\hbar^2 k^2}{2m}+2gn\right)}.
    \label{eq:bog}
\end{equation}
Now at low-$k$ there are collective excitations with a sound-like form $\epsilon/\hbar= c_0 k$, where $c_0=\sqrt{gn/m}$.
This is crucial for superfluidity in Bose condensed systems, as it ensures (unlike in an ideal gas) a nonzero superfluid critical velocity (according to the Landau criterion).
The sound-like excitation spectrum was first seen in liquid $^{4}$He~\cite{Nozieres:1990}. In ultracold atomic gases, following initial studies of collective excitations ~\cite{Jin:1996a,Mewes:1996b,Andrews:1997}, a direct comparison to Eq.~(\ref{eq:bog}) was made possible by using Bragg spectroscopy on an ultracold Bose--Einstein condensate [see \cite{Kozuma:1999a,Stenger:1999b,Steinhauer:2002} and Fig.~\ref{fig:Sound}(a)].

At finite temperature the situation is further complicated by the presence of thermal excitations, which lead to two distinct sound modes, the (faster) first and (slower) second sound. The two modes can be understood, in the hydrodynamic limit, within the two-fluid model\cite{Landau:1941}, which separately considers the normal and superfluid components. The nature of the sound modes changes depending on how compressible the Bose fluid is. 
In the almost incompressible liquid helium, first and second sound are, respectively, density and temperature (or entropy) waves. Instead, in a dilute Bose fluid (which is highly compressible) the two modes are excitations of mainly the normal and superfluid components [see~\cite{Hilker:2022} and Fig.~\ref{fig:Sound}(b)].

Quantised vortices are another important type of excitation that feature in interacting Bose-condensed gases.
Such vortices carry angular momentum quantised in units of $\hbar/m$ and have vanishing density along the vortex core over a healing length $1/\sqrt{8\pi n a}$ [see \cite{Madison:2000a} and Fig.~\ref{fig:Sound}(c)].
In a system containing many vortices the interactions between them can result in the formation of vortex lattices [see \cite{AboShaeer:2001} and Fig.~\ref{fig:Sound}(d)]. Note that these  lattices are closely related to those that occur in type-II superconductors in the presence of a magnetic field; rotation can be thought of as an artificial magnetic field, which is an example of a synthetic gauge field (see \emph{e.g.}~\cite{Goldman:2014,Zhai:2015,Aidelsburger:2018}).

\subsection{Quantum depletion}
\begin{figure}[t!]
\centering
\centerline{\includegraphics[width=\columnwidth]{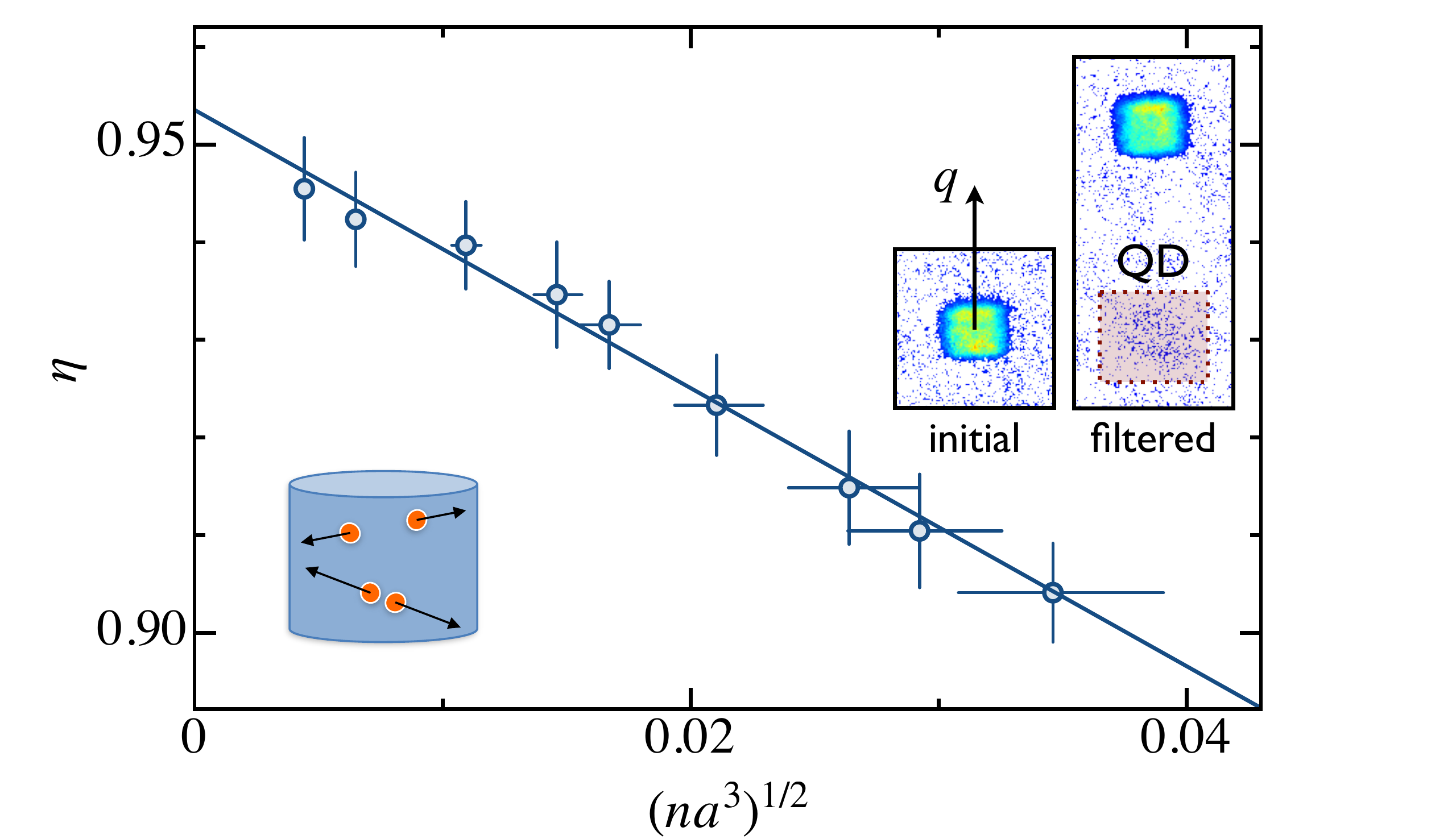}}
\caption{$\bf{\color{myColor}}$ \textbf{Measurement of quantum depletion.}
The maximal diffracted fraction $\eta$ versus the interaction parameter $({na^3})^{1/2}$; here a Bragg filtering technique was used to spatially separate the condensate from the high-$k$ components of the gas (including the quantum depletion; see top inset).
Assuming both perfect $k$-space separation and filtering, $\eta$ corresponds to the condensed fraction.
A linear fit (solid line) gives a slope $\gamma=1.5(2)$, quantitatively confirming the Bogoliubov prediction $\gamma=8/(3\sqrt{\pi})\approx 1.5$.
The fact that the intercept is only close to unity, arises from an imperfect $k$-space separation and a non-zero initial temperature~\cite{Lopes:2017b}.
The cartoon (bottom inset) depicts the coherent excitations out of the (blue) condensate, which occur as pairs of atoms with opposite momenta.
Figure adapted from \cite{Lopes:2017b}.
}
\label{fig:QD}
\end{figure}

In addition to modifying the excitation spectrum, interparticle interactions also modify the ground state, an effect known as quantum depletion. This entails the coherent expulsion of particles from the single particle ($p=0$) ground state, even at $T=0$; in essence, it is energetically favourable for some higher momentum states to be occupied.

The most striking example of quantum depletion occurs in liquid $^{4}$He, where even though at $T=0$ the system is 100\% superfluid, the condensed fraction $n_0/n$ is only around 10\%. However, liquid $^{4}$He is a strongly interacting fluid, and so theoretical predictions are challenging.

Instead, for a weakly interacting gas [$(na^3)^{1/2} \ll 1$], a theoretical description is more tractable, with \mbox{$T=0$} Bogoliubov theory predicting~\cite{Lee:1957b}
\begin{equation}
\frac{n_0}{n}=1-\frac{8}{3\sqrt{\pi}}(na^3)^{1/2}\,.  
\label{eq:qd}
\end{equation}
This famous prediction was experimentally verified in a box-trapped (homogeneous) ultracold gas~(see \cite{Lopes:2017b} and Fig.~\ref{fig:QD}).
Within the same theoretical framework, the ground-state energy is also modified, as first measured in~\cite{Navon:2011}.

\subsection{Interaction effects near $\Tc$}

\begin{figure}[t!]
\centering
\includegraphics[width=1\columnwidth]{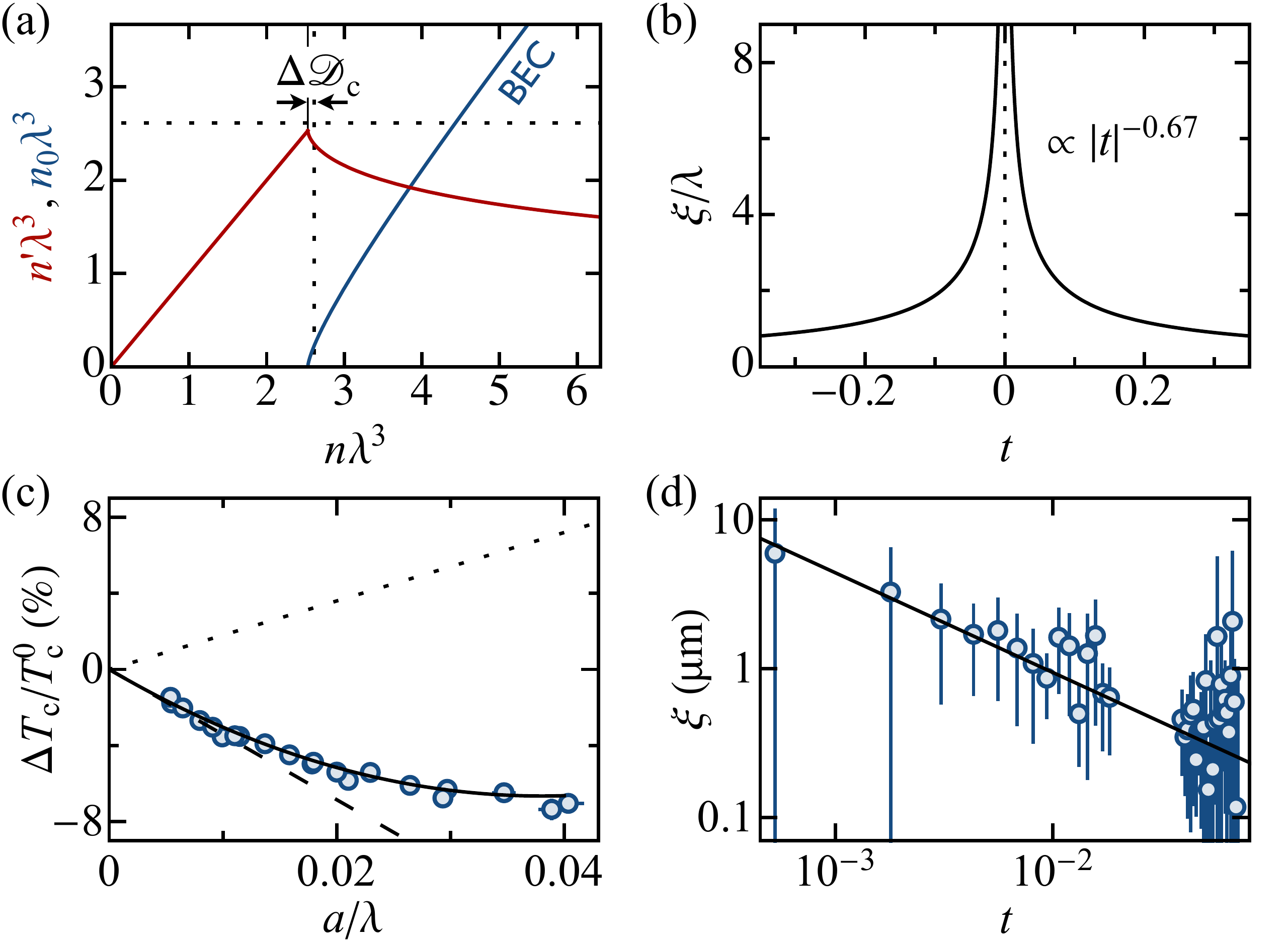}
\caption{\textbf{Thermodynamics near $\Tc$.}
In (a,b) we illustrate the effects of interactions near $\Tc$ in a homogeneous system, while (c,d) highlight related experimental results.
(a) Excited- and ground-state phase-space densities as a function of the total phase-space density for an interacting Bose gas [\emph{cf.} Fig.~\ref{fig:idealgas}(b)]. For $n\lambda^3>\mathcal{D}_{\rm c}$, $n'\lambda^3$ is no longer saturated at $\mathcal{D}_{\rm c}$, but decreases as $n\lambda^3$ increases.  The critical phase-space density also undergoes a small shift. (b) The correlation length $\xi$ versus reduced temperature $t=(T-\Tc)/\Tc$ near the BEC transition.
The interacting Bose gas is in the 3D XY Model universality class, which has $\xi \sim t^{-\nu}$ with $\nu \approx 0.67$. (c) Measured $\Tc$ shift versus interaction strength for a harmonically trapped Bose gas \cite{Smith:2011}; an upward shift from the predominant MF geometric effect (dashed line) is seen, but quantitatively comparing this to the homogeneous system result (dotted line) was not possible. (d) Measurement of the correlation length critical exponent in a ultracold Bose gas \cite{Donner:2007}; the line is a fit to the data, which gives $\nu=0.67(13)$ consistent with the 3D XY Model.
Panels adapted from: (c)\cite{Smith:2011}, (d)\cite{Donner:2007}.
}
\label{fig:thermo}
\end{figure}

We now turn to the effect of interactions on the thermodynamics of a Bose condensed gas, focusing on the behavior close to $\Tc$; these effects are summarized in Figs.~\ref{fig:thermo}(a,b).

First, in the (partially) condensed phase, interactions cause $n'\lambda^3$ to no longer be saturated at $\mathcal{D}_{\rm c}$  [\emph{cf.} Figs. \ref{fig:idealgas}(b) and \ref{fig:thermo}(a)], but instead it decreases for increasing $n\lambda^3$.
This effect comes about (at the MF level) because an atom can reduce its interaction energy by entering the condensate. 
Note that for a harmonically trapped gas the opposite effect occurs, with the excited-state population increasing as the system goes more deeply into the condensed state \cite{Tammuz:2011}.

Secondly, condensation occurs at $\mathcal{D}_{\rm c}<\mathcal{D}_{\rm c}^0$ (or equivalently for $\Tc>\Tc^{0}$ at fixed $n$); here the superscript~$^0$ denotes the ideal-gas results. 
This effect can only be understood at the BMF level, as long-wavelength fluctuations become increasingly important close to $\Tc$. Calculating the shift was a significant theoretical challenge and it took several decades to reach a consensus (see \cite{Andersen:2004, Arnold:2001, Baym:2001, Holzmann:2004} for a review); the shift is predicted to be \cite{Arnold:2001, Kashurnikov:2001}
\begin{equation}
    \frac{\Delta \Tc}{\Tc^0} \approx 1.8 \frac{a}{\lambda},
    \label{eq:tcshift}
\end{equation}
but this result has yet to be confirmed experimentally. 

In a harmonic trap, a relatively simple geometric (MF) effect results in a negative $\Tc$ shift, with the more interesting BMF effects expected to be pushed to higher order in $a/\lambda$. As shown in Fig.~\ref{fig:thermo}(c), measurements in a harmonic trap \cite{Smith:2011} did show an upward shift from the MF prediction (dashed line), but making a quantitative comparison with Eq.~(\ref{eq:tcshift}) (dotted line) was not possible.   

Thirdly, interactions modify the critical exponents of the BEC transition, changing the universality class to that of the 3D XY model.
In particular, the correlation length critical exponent is predicted to change from $\nu=1$ to $\nu \approx 0.67$~\cite{Burovski:2006b}; see Fig.~\ref{fig:thermo}(b). 
This exponent has been measured by locally probing a harmonically trapped ultracold gas [see \cite{Donner:2007} and Fig.\,\ref{fig:thermo}(d)], and also with much higher precision in ${^4}$He~(see \cite{Burovski:2006}).

\section{Unitary contact interactions}
\label{sec:Unitary}

\begin{figure}[b!]
\centering
\centerline{\includegraphics[width=\columnwidth]{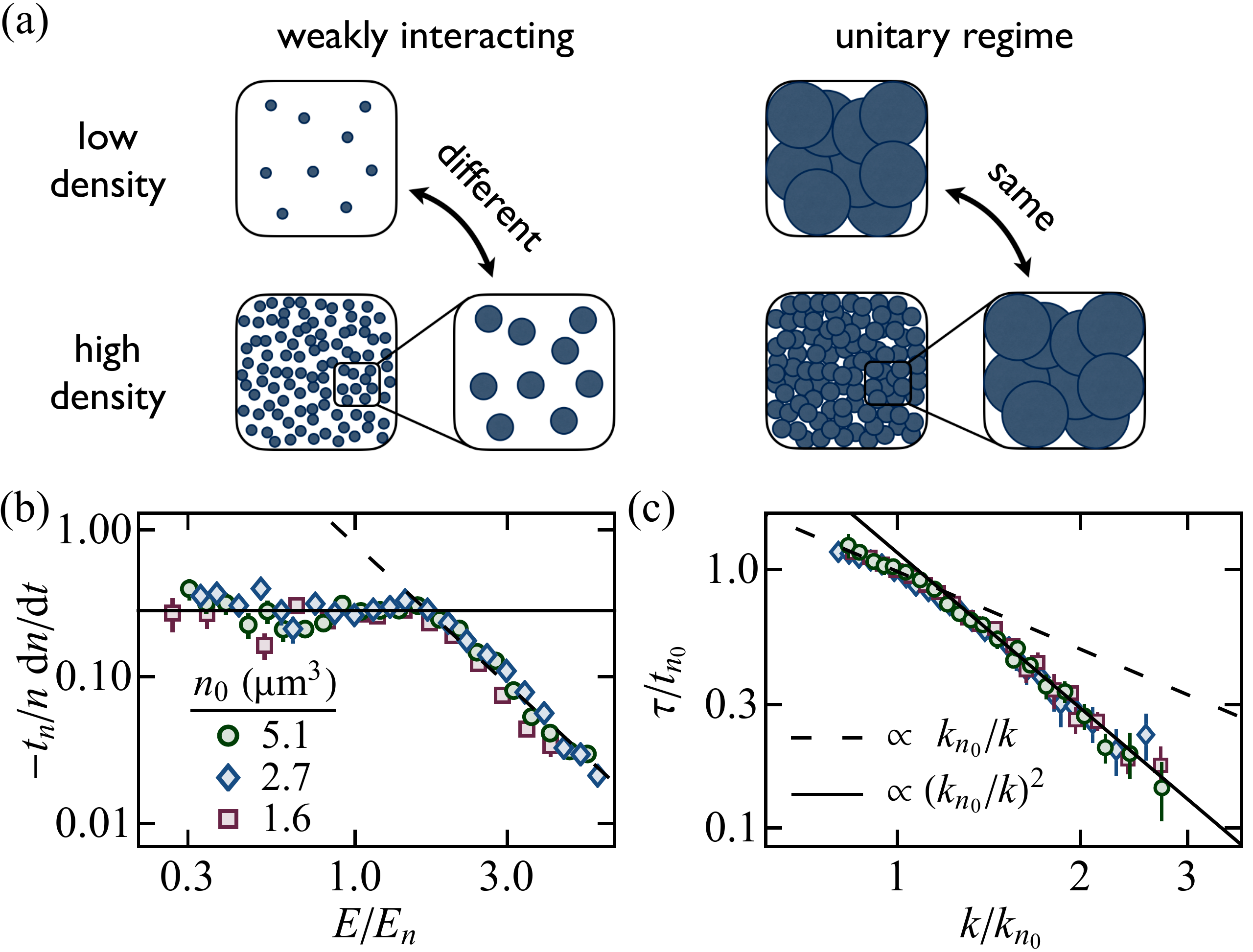}}
\caption{\textbf{Universal behavior in unitary Bose gases.} (a) In a weakly interacting Bose gas (left) the interactions, characterized by $a$, set the relevant lengthscale (size of the circles in the cartoon), so that when `zooming in' on a high-density gas it looks different to a low-density gas. In the unitary regime (right), upon zooming in, the high-density gas instead looks the same as the low-density one, exhibiting scale invariance.
(b,c)~Examples of experimental verification of the universality hypothesis \cite{Eigen:2017,Eigen:2018}. For initially degenerate Bose gases of initial density $n_0$ quenched to unitarity, both (b) the dimensionless loss rate versus $E/E_n$ \cite{Eigen:2017} and (c) the momentum-dependent prethermal relaxation timescale $\tau/ t_{n_0}$ \cite{Eigen:2018} are universal functions that depend solely on the density.
Panels adapted from: (b)\cite{Eigen:2017}, (c)\cite{Eigen:2018}.
}
\label{fig:unin}
\end{figure}

Here we consider Bose-condensed systems in the strongly interacting limit ($na^3\gg1$), experimentally accessible at a Feshbach resonance, where $a$ diverges. In this so-called unitary regime, the interactions between particles are as strong as allowed by quantum mechanics. The fact that the value of the diverging $a$ can no longer matter leads to the universality hypothesis, in which the interparticle spacing $n^{-1/3}$ is the only relevant lengthscale for a degenerate gas [see \cite{Cowell:2002,Ho:2004a} and Fig.~\ref{fig:unin}(a)]. This also sets the natural momentum, energy, and time scales:
\begin{equation}
\hbar k_n = \hbar (6\pi^2 n)^{1/3},~ E_n = \hbar^2 k_n^2/(2m),~{\rm and}~t_n = \hbar/E_n\,.
\end{equation}

In contrast to their widely studied Fermi counterparts\cite{Inguscio:2007, Zwerger:2011,Zwierlein:2014}, unitary Bose gases feature dramatically enhanced particle loss
\footnote{For $i$-body loss, dimensional analysis gives $\dot{n}/n\propto \hbar/m a^{3i-5}n^{i-1}$~\cite{Thoegersen:2008,Mehta:2009}, which for $na^3\gg1$ gives $\dot{n}/n\propto 1/t_n$ for all $i$. Also note that in Bose gases Efimov physics further modulates the atom loss~\cite{Efimov:1970,Braaten:2007,Naidon:2017}.}
and associated heating, which makes their study an inherently dynamical problem and raised the question whether such a gas can ever exist in equilibrium.

The experimental approach has been to forsake equilibrium and quench into the unitary regime (by rapidly increasing $a$), to then observe the ensuing dynamics.
These experiments have shown a remarkable degree of universality [see~\cite{Makotyn:2014,Klauss:2017,Eigen:2017,Eigen:2018} and Fig.~\ref{fig:unin}(b,c)].
The post-quench dynamics also revealed that the gas, even though it is not in true thermal equilibrium, does attain a quasi-equilibrium (prethermal) state~\cite{Berges:2004,Yin:2016,Makotyn:2014,Eigen:2018}. Eventually, the gas decays and heats, ultimately following the dimensionless loss rate for a thermal unitary gas $-t_n\dot{n}/n\propto (E_n/E)^2$, where $E$ is the kinetic energy per particle (dashed line in Fig.~\ref{fig:unin}(b); see also thermal unitary-gas measurements~\cite{Rem:2013,Fletcher:2013,Eismann:2016}).

Intriguingly, the observed post-quench prethermal state featured: (1) a momentum-dependent relaxation timescale approximately given by $\hbar/\epsilon(k)$ [see Eq.~(\ref{eq:bog})] with $gn$ replaced by $E_n$ [see Fig.~\ref{fig:unin}(c)], and (2) a universal momentum distribution, which implies a non-zero condensed fraction. This hints at the possibility of a novel superfluid state consisting of Efimov trimers \footnote{The Efimov effect is a quantum three-body effect first discussed in nuclear physics in 1970~\cite{Efimov:1970}}, which has also been theoretically predicted~\cite{Piatecki:2014,Musolino:2022}.

\section{Attractive contact interactions and dipolar interactions} 
\label{sec:AttractiveAndDipolar}

So far, we have only discussed the effect of repulsive contact interactions on a single component Bose gas, which for $na^3 \ll 1$ leads to relatively modest changes (see Section~\ref{sec:WeakRepulsiveContact}).   

We now turn to scenarios where the interactions are (at least partially) attractive. This can lead to far more dramatic consequences and also to situations where BMF effects play a critical role. In the following, we in turn discuss: single component condensates with weak attractive contact interactions, quantum mixtures, and situations in which the atoms interact via both contact and dipolar interactions.

\subsection{Attractive contact interactions}

For attractive contact interactions ($a<0$) in the thermodynamic limit, Bose--Einstein condensates are not stable, and are susceptible to collapse. 
This can be seen by examining the Bogoliubov excitation spectrum [Eq.~(\ref{eq:bog})] and noting that negative values of $g$ lead to imaginary excitation energies at low $k$, signifying unstable (exponentially growing) modes; this is sometimes referred to as a phonon instability.

Trapped (finite-sized) 3D condensates can be stabilized by the ground-state kinetic energy (quantum pressure), such that for a given atom number $N$, collapse only occurs above a critical attractive $|\ac|$. Note that in lower-dimensional systems the situation is somewhat different, \emph{e.g.}~in 1D the balance of kinetic and interaction energy permits the formation of solitons\cite{Strecker:2002,Khaykovich:2002}.

\begin{figure}[t!]
\centering
\includegraphics[width=1\columnwidth]{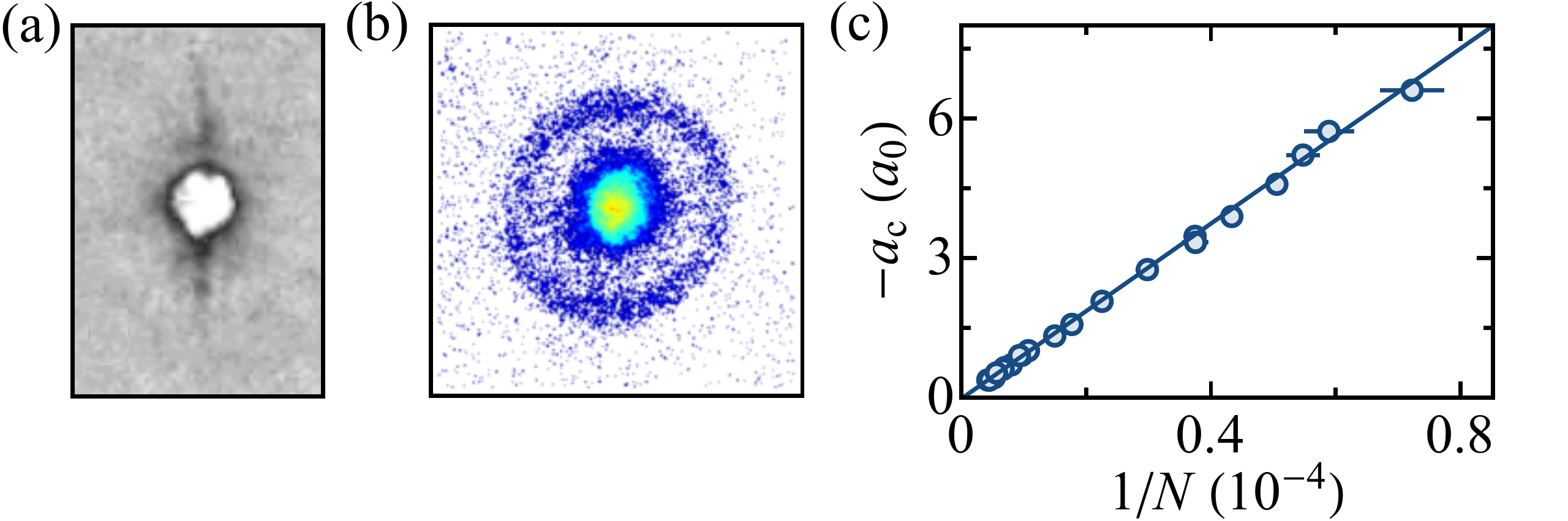}
\caption{
\textbf{Condensate collapse.}
(a,b) Absorption images of ultracold quantum gases following wave collapse for (a) a harmonically trapped~\cite{Donley:2001} and (b) a box-trapped~\cite{Eigen:2016} cloud. (c) Measured critical $\ac$ (in units of the Bohr radius $a_0$) versus $1/N$ for collapse of a condensate confined in a cylindrical box trap of size $L$ (length equal to diameter)\cite{Eigen:2016}. The solid line shows a fit to the data, which gives $-\ac= 0.16(4)L/N$ consistent with the Gross--Pitaevskii equation prediction $-\ac= 0.17L/N$.
Panels adapted from: (a)\cite{Donley:2001}, (b,c)\cite{Eigen:2016}.
}
\label{fig:collapse}
\end{figure}

Condensate collapse was originally explored for atoms confined in harmonic traps~\cite{Gerton:2000,Roberts:2001,Donley:2001} and later also for those in 3D box potentials~\cite{Eigen:2016}. Figures~\ref{fig:collapse}(a,b) show, respectively, experimental images of the collapse in harmonically and box-trapped gases.
In both cases, the critical point is given by $N \ac \propto L$, where $L$ is the linear size of the single-particle ground state [see \cite{Gerton:2000,Roberts:2001,Donley:2001,Eigen:2016} and Fig.~\ref{fig:collapse}(c)]. 
The basic picture can again be understood from Eq.~(\ref{eq:bog}): Equating $2gn$ to $\hbar^2 k^2/(2m)$ with (the minimum) $k\sim \pi/L$ (using $n\approx N/L^3$) predicts a critical value of $N\ac/L\approx-0.2$ [\emph{cf.}~Fig.~\ref{fig:collapse}(c)].
More quantitatively, the collapse of such matter waves can be described using the Gross--Pitaevskii equation.

In single-component condensates with attractive contact interactions, the BMF effects are typically negligible as $a$ is close to zero. 
However, in situations where two (potentially) large MF effects cancel, BMF effects can become important and even lead to qualitatively new effects; examples of such scenarios are explored in the two following sections.

\subsection{Interacting Bose mixtures} 

\begin{figure}[t!]
\centering
\includegraphics[width=1\columnwidth]{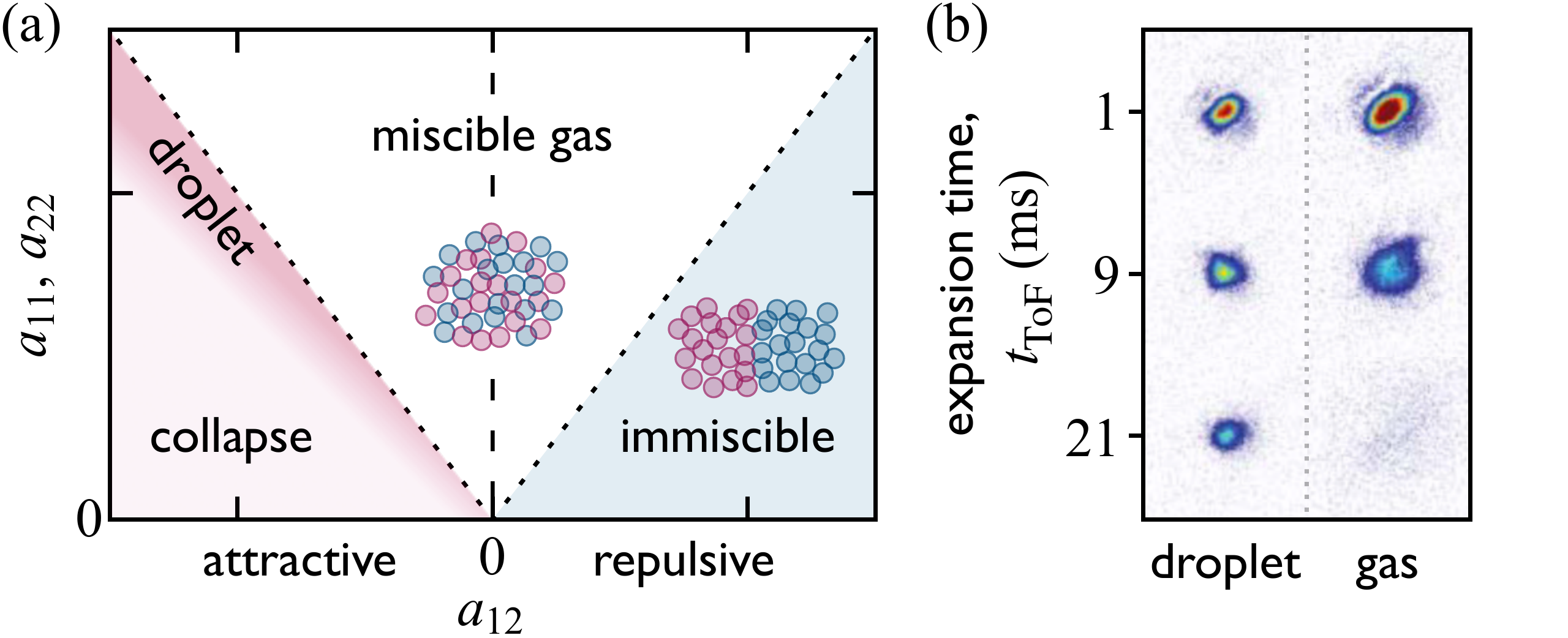}
\caption{\textbf{Realising quantum droplets in Bose mixtures.} 
(a) Illustrative phase diagram for $a_{11}=a_{22}>0$ showing the miscible, immiscible, collapse, and droplet regimes. At the MF level for a homogeneous gas, the miscible region is given by $-\sqrt{a_{11}a_{22}} <a_{12}<\sqrt{a_{11}a_{22}}$ (between the dotted lines). Quantum droplets can be formed in a sliver of the phase diagram near $\delta a = a_{12} + \sqrt{a_{11} a_{22}} =0$, where BMF effects can become important (see text).
(b) Images demonstrating the existence of a droplet phase\cite{Cabrera:2018}; for $\delta a=-3.2 a_0$ (left) the cloud remains self-bound after release from the trap, compared to $\delta a=1.2 a_0$ (right), which shows the typical expansion of a miscible gas. 
Panel (b) adapted from \cite{Cabrera:2018}.
}
\label{fig:mix}
\end{figure}

In a two-component quantum mixture with contact interactions between the constituents, there are three relevant scattering lengths: two intrastate ones ($a_{11}$ and $a_{22}$), and an interstate one ($a_{12}$). 
A sketch of the zero-temperature phase diagram of such a mixture is shown in Fig.~\ref{fig:mix}(a) for the case $a_{11}=a_{22} > 0$. 

At the MF level, for $a_{12}>\sqrt{a_{11}a_{22}}$ the interstate repulsion causes the two components to separate (the system is immiscible), whereas for $a_{12}<-\sqrt{a_{11}a_{22}}$ the interstate attraction overwhelms the intrastate repulsion, leading to collapse. In between, the two components are miscible - forming overlapping condensates.
The miscible-immiscible transition has been studied since the early days of atomic-gas condensates (see \emph{e.g.} \cite{Hall:1998b,Stenger:1998b,Papp:2008b}).

At the collapse boundary, where the MF terms (almost) cancel, the residual mean-field interaction, $\delta a=a_{12} + \sqrt{a_{11}a_{22}}$, is small. However, in contrast to a single-component gas near the collapse boundary, the BMF effects are not necessarily negligible. The BMF effects are typically repulsive in nature and scale more strongly with density than the MF effects that drive the collapse, opening the possibility of quantum liquid droplets where (at an appropriate density) the attractive MF attraction can be balanced by BMF repulsion~\cite{Petrov:2015}. Such droplets were realised in 2018 [see \cite{Cabrera:2018, Semeghini:2018} and Fig.~\ref{fig:mix}(b)]. The `liquid' label refers to the fact that, in principle, the droplets are stabilized at a fixed density and are self bound; in practise, the high densities involved lead to (three-body) losses that limit the droplet lifetime.
In the special case where $\delta a \approx 0$, a so-called `LHY fluid' is formed, which only features BMF interactions\cite{Skov:2021}.

\subsection{Dipolar interactions} 
\label{sec:Dipolar}

In this section we briefly review novel effects of dipole-dipole interactions on Bose-condensed gases (see \emph{e.g.}~\cite{Lahaye:2009, Norcia:2021b, Chomaz:2022} for a more detailed treatment). Such interactions, which arise for particles with a permanent dipole moment, are long-range and anisotropic; the interaction potential for two polarised dipoles a distance $r=|\bf{r}|$ apart is given by
\begin{equation}
    V_{\rm dd}=\frac{C_{\rm dd}}{4 \pi} \frac{1-3 \cos^2  \theta}{r^3}\,,
\end{equation}
where $\theta$ is the angle between $\bf r$ and the polarization axis, and $C_{\rm dd}$ the dipole coupling constant, which is given by $\mu_0 \mu^2$ ($d^2/\epsilon_0$) for magnetic (electric) dipoles. The strength of dipolar interactions are also commonly captured by the dipolar length $a_{ \rm dd} = C_{\rm dd} m/(12 \pi \hbar^2)$.

For a homogeneous gas with both contact and dipole-dipole interactions, the Bogoliubov excitation spectrum [Eq.~(\ref{eq:bog})] is modified such that 
\begin{equation}
    \epsilon({\bf k})=\sqrt{\frac{\hbar^2 k^2}{2m}\left(\frac{\hbar^2 k^2}{2m}+2gn+2C_{\rm dd}\left(\cos^2\alpha-\tfrac{1}{3}\right)n \right)}\,,
    \label{eq:bog_d}
\end{equation}
where $\alpha$ is the angle between $\bf k$ and the polarization direction.
Several consequences of dipolar interactions can directly be inferred from Eq.~(\ref{eq:bog_d}). 

\begin{figure}[b!]
\centering
\includegraphics[width=1\columnwidth]{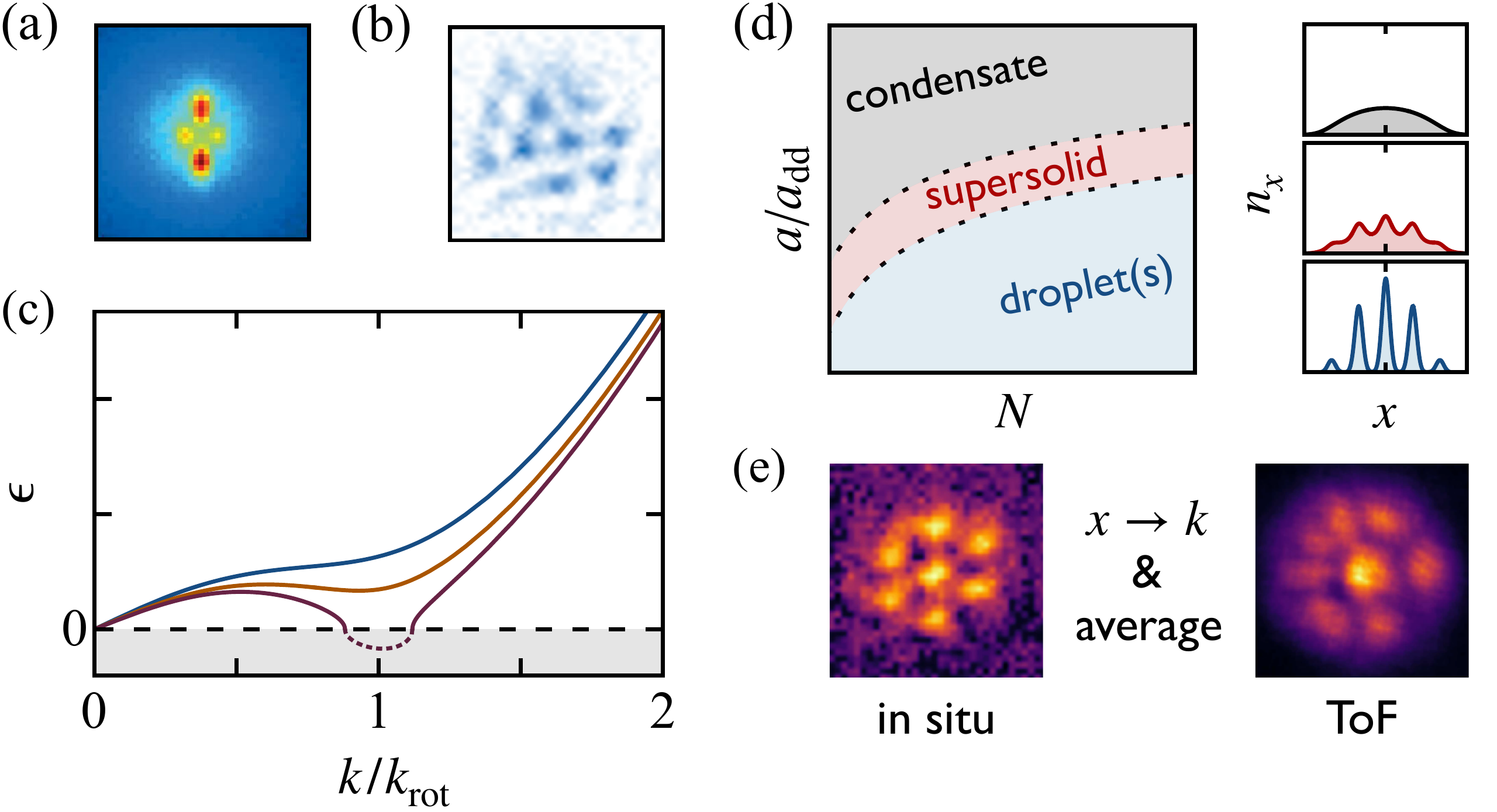}
\caption{\textbf{Dipolar Bose gases.} 
(a) Absorption image revealing the effects of dipolar interactions following the collapse and subsequent explosion of a condensate\cite{Lahaye:2008}. 
(b) Observation of dipolar quantum droplets\cite{Kadau:2016}.  
(c) Sketch of the roton excitation spectrum seen in dipolar gases that are confined along their polarisation direction. As the relative strength of dipole-dipole interactions increases the roton minimum deepens and then triggers (red line) an instability near $k_{\rm rot}$. 
(d) Cartoon of the phase diagram of a dipolar gas in a cigar-shaped trap. Starting with a condensate, reducing $a/a_{\rm dd}$ triggers the roton instability resulting in an array of droplets. For a sliver of the phase diagram a supersolid state exists, in which the droplets are phase coherent \cite{Tanzi:2019,Bottcher:2019,Chomaz:2019}.
(e) Observation of a 2D supersolid \cite{Norcia:2021, Bland:2022}; (left) an in-situ image revealing a hexagonal density modulation and (right) a time-of-flight (ToF) image, averaged over multiple realizations, which demonstrates the global phase coherence of the system. 
Panels adapted from: (a) \cite{Lahaye:2008}, (b) \cite{Kadau:2016}, (e) \cite{Bland:2022}.
}
\label{fig:dip}
\end{figure}

Even for weak dipolar interactions the excitation spectrum is anisotropic, which leads to anisotropic sound waves~\cite{Bismut:2012}. Moreover, for $C_{\rm dd}/3>g$ (or equivalently $a_{\rm dd}>a$) unstable phonon modes exist and (at least from a MF perspective) the gas can undergo collapse, with the angular dependence of the interactions leading to spectacular collapse dynamics [see \cite{Lahaye:2007} and Fig.~\ref{fig:dip}(a)].
As in the case of a Bose mixture, collapse occurs at a nonzero $g$, opening up the possibility of the collapse being arrested by BMF effects, which again can result in the formation of quantum droplets~[see \cite{Kadau:2016,Schmitt:2016,Ferrier-Barbut:2016,Chomaz:2016} and Fig.~\ref{fig:dip}(b)]. 

The presence of a trap tends to stabilise the system.
A particularly interesting situation arises when a dipolar gas is tightly confined along the polarisation direction, which leads to the development of an in-plane roton-like excitation spectrum [see \cite{Petter:2019} and Fig.~\ref{fig:dip}(c)]. Note that although similar to the roton excitation spectrum originally seen in liquid $^{4}$He~(see \emph{e.g.}\cite{Donnelly:1981}), its microscopic origin is different, occurring as the consequence of both confinement and long-range anisotropic interactions.
When the roton gap closes, an instability occurs at finite $k=k_{\rm rot}$, such that without any stabilisation mechanism a runaway density modulation would ensue. 
However, the collapse can yet again be halted by BMF effects, leading to the formation of a droplet array. For a narrow range of parameters close the stability boundary, these droplets remain phase coherent, signalling the existence of a supersolid state -- a state with both long-range phase order (and hence superfluidity) and long-range spatial order (the density modulation) [see \cite{Tanzi:2019,Bottcher:2019,Chomaz:2019,Norcia:2021,Bland:2022} and Fig.~\ref{fig:dip}(d,e)].

\section{Summary and Outlook}
\label{sec:conclusions}
\vspace{-2em}
In summary, we have reviewed effects of both contact and dipolar interactions on Bose-condensed gases. Many of the observed phenomena have been long predicted, if only more recently confirmed in ultracold gases, while others, such as the observation of supersolidity, were largely unexpected. Looking forward, further breakthroughs in both categories are within reach. A major longstanding task is to experimentally resolve the simple question of how weak interactions shift the BEC transition temperature.  
On the more exploratory side, the realms of strong contact and dipolar interactions offer the possibility of novel many-body physics.

\section{Acknowledgments}

We thank Zoran Hadzibabic for comments on the manuscript. C.~E. acknowledges support from Jesus College (Cambridge).

\end{document}